\documentclass{article}
\usepackage{spconf,amsmath,graphicx}
\usepackage{amsmath,amsfonts,amssymb,pifont}
\usepackage{adjustbox}
\usepackage{comment}
\usepackage{xparse}
\NewDocumentCommand\emojiclap{}{\includegraphics[scale=0.07]{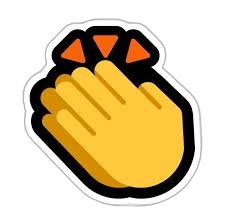}}
\usepackage{multirow,makecell}

\newcommand{\cmark}{\text{\ding{51}}}
\newcommand{\xmark}{\text{\ding{55}}}

\title{CLAP \emojiclap: Learning Audio Concepts From Natural Language Supervision \vspace{-7pt}}
\name{Benjamin Elizalde, Soham Deshmukh, Mahmoud Al Ismail, Huaming Wang \vspace{-8pt}}
\address{Microsoft \\
\{benjaminm, sdeshmukh, malismail, huawang\}@microsoft.com \vspace{-8pt}}

\begin{document}
%
\maketitle
\begin{abstract}
\vspace{-0.01in}
Mainstream Audio Analytics models are trained to learn under the paradigm of one class label to many recordings focusing on one task. Learning under such restricted supervision limits the flexibility of models because they require labeled audio for training and can only predict the predefined categories. Instead, we propose to learn audio concepts from natural language supervision. We call our approach Contrastive Language-Audio Pretraining (CLAP), which learns to connect language and audio by using two encoders and a contrastive learning to bring audio and text descriptions into a joint multimodal space. We trained CLAP with 128k audio and text pairs and evaluated it on 16 downstream tasks across 8 domains, such as Sound Event Classification, Music tasks, and Speech-related tasks. Although CLAP was trained with significantly less pairs than similar computer vision models, it establishes SoTA for Zero-Shot performance. Additionally, we evaluated CLAP in a supervised learning setup and achieve SoTA in 5 tasks. Hence, CLAP's Zero-Shot capability removes the need of training with class labels, enables flexible class prediction at inference time, and generalizes to multiple downstream tasks.
\end{abstract}
\vspace{-0.15in}
\section{Introduction}\label{sec:intro}
\vspace{-0.05in}
\begin{figure*}[ht]
   \centering
     \includegraphics[width=\textwidth,scale=0.9]{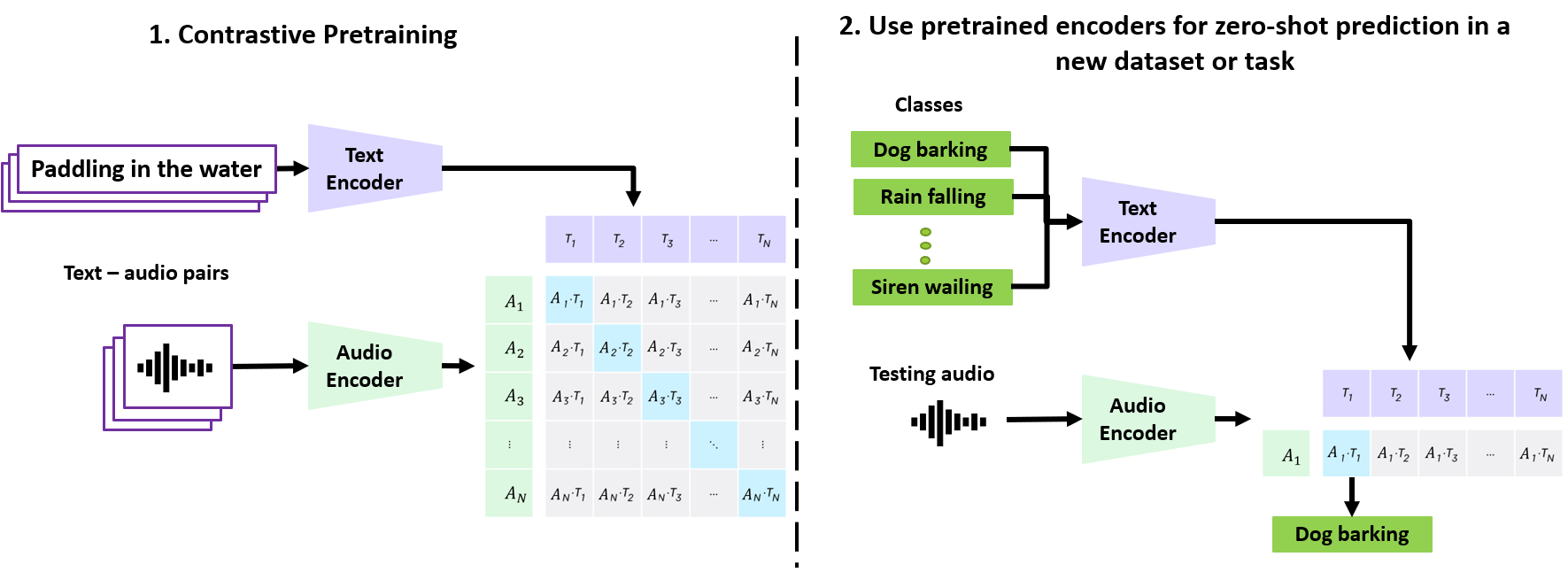}
     \caption{CLAP \emojiclap jointly trains an audio and a text encoder to learn the (dis)similarity of audio and text pairs in a batch using contrastive learning. At testing time, the pretrained encoders are used to extract audio embeddings from the testing audio and text embeddings from the class labels. Zero-Shot linear classification is achieved by computing cosine similarity between the embeddings. \vspace{-0.08in}}
     \label{fig:clap_diagram}
\end{figure*}

The human auditory system can hear sounds and extract the kind of decisions or meanings we need to interact with our surroundings~\cite{lyon2010machine}. For example, if we are in a soccer game and suddenly hear the crowd cheering joyfully, we can assume the local team scored! Computer models aim to understand audio cues by automatically process audio signals and extract meaning~\cite{lyon2010machine}. Mainstream Machine Learning models break the human hearing into tasks, such as the classification of sound events and acoustic scenes. Such models are trained by associating audio recordings to class labels of predefined categories for a specific task and can only predict specific categories~\cite{mesaros2019sound}. Learning under such restricted supervision limits the flexibility to predict unseen classes. 

Self-Supervised Learning (SSL) pretrains models with unlabeled audio, avoiding the limited supervision of learning from class labels. However, SSL excludes semantic knowledge from natural language. The pretrained model is then adapted to a downstream task in a supervised setup learning under the class label paradigm ~\cite{zhang2021bigssl, chen2021wavlm, baevski2020wav2vec}. Mainstream and SSL models have static output layers that can only predict the predefined categories. On the other hand, models that enable Zero-Shot predictions can take an input audio and yield a prediction score for any class typed by the user. Zero-shot requires no training stage so there are no predefined categories. To enable such flexibility and generalization, models need to learn the relationships between the acoustic semantics and language semantics.

A middle path between both approaches is to learn audio concepts from natural language supervision, which is under-explored. Computer Vision has successfully developed models that learn image representations with natural language supervision, achieving high performance across different downstream tasks in Zero-Shot predictions and adapted (e.g. finetuned) to target datasets in a supervised setup. Examples are Open AI's CLIP~\cite{radford2021learning}, Florence~\cite{yuan2021florence}, and~\cite{jia2021scaling}. In the audio domain, Wav2clip~\cite{wav2clip} and Audioclip~\cite{audioclip} distill from CLIP and are trained with audio and class labels from AudioSet~\cite{audioSet} instead of audio and natural language. Although the performance was promising, it is yet to be explored how natural language can benefit flexibility and generalization to new classes and tasks.

We call our approach Contrastive Language-Audio Pretraining (CLAP), which learns to connect natural language and audio by using two encoders and contrastive learning to bring audio and text descriptions into a joint multimodal space. Our main contributions are first, to introduce our CLAP model trained with 128k audio and text pairs. Second, our model enables Zero-Shot predictions, thus it removes the need of training and forcing a predefined set of categories and enables flexible class prediction at inference time. Third, CLAP generalizes to 16 downstream tasks across 8 domains by establishing Zero-Shot SoTA performance. We also include insights about the model in Section~\ref{sec:results}.
\vspace{-0.15in}

\section{Method}\label{sec:method}
\vspace{-0.05in}
CLAP is illustrated in Fig~\ref{fig:clap_diagram}. The input is audio and text pairs passed to an audio encoder and a text encoder. Both representations are connected in joint multimodal space with linear projections. The space is learned with the (dis)similarity of audio and text pairs in a batch using contrastive learning. The pretrained encoders with their projection layers can be used to compute audio and text embeddings and enable Zero-Shot Classification. Our method is inspired by the CLIP model~\cite{radford2021learning}.

\subsection{Contrastive Language-Audio Pretraining}
\vspace{-0.05in}

Let the processed audio be $X_a$ s.t. $X_a \in \mathbb{R}^{F \times T}$ where $F$ are the number of spectral components (e.g. Mel bins) and $T$ are the number of time bins. Let the text be represented by $X_t$. Each audio-text pair in a batch of $N$ is represented as $\{X_a, X_t\}_i$ where $i \in [0,N]$. For convenience, we dropped the $i$ notation, and henceforth $\{X_a, X_t\}$ will denote a batch of N. 

From the pairs, the audio and text are passed through an audio encoder and a text encoder respectively. Let $f_a(.)$ represent the audio encoder and $f_t(.)$ represent the text encoder. For a batch of N:
\begin{equation}
    \hat{X}_a = f_a(X_a); 
    \hat{X}_t = f_t(X_t)
\end{equation}
where $\hat{X}_a \in \mathbb{R}^{N \times V}$ are the audio representations of dimensionality $V$, and $\hat{X}_t \in \mathbb{R}^{N \times U}$ are the text representations of dimensionality $U$. 

We brought audio and text representations, $\hat{X}_a$ and $\hat{X}_t$, into a joint multimodal space of dimension $d$ by using a learnable linear projection: 
\begin{equation}
    E_a = L_a(X_a);
    E_t = L_t(X_t)
\end{equation}
where $E_a \in \mathbb{R}^{N \times d}$, $E_t \in \mathbb{R}^{N \times d}$, $L_a$ and $L_t$ are the linear projections for audio and text respectively. 

Now that the audio and text embeddings ($E_a$, $E_t$) are comparable, we can measure similarity:
\begin{equation}
    C = \tau*(E_t \cdot E_a^\top)
\end{equation}
where $\tau$ is a temperature parameter to scale the range of logits. The similarity matrix $C \in \mathbb{R}^{N \times N}$ has $N$ correct pairs in the diagonal and $N^2-N$ incorrect pairs in the off-diagonal. 
\begin{equation}
     \mathcal{L} = 0.5 * (\ell_{text}(C) + \ell_{audio}(C))
\end{equation}
where $\ell_{k} = \frac{1}{N}\sum_{i=0}^{N} \log diag (softmax(C))$ along text and audio axis respectively. We used this symmetric cross-entropy loss ($\mathcal{L}$) over the similarity matrix to jointly train the audio encoder and the text encoder along with their linear projections. 

\vspace{-0.05in}
\subsection{Zero-Shot Linear Classification}
\label{sec:method,subsec:zeroshot} 
\vspace{-0.05in}

For Zero-Shot classification, we used CLAP's ability to determine the similarity between audio and text. Let's consider a target dataset with $C$ class labels and $N$ test audios. First, we compute audio embeddings and text embeddings for $N$ audios and $C$ classes using the pretrained encoders and their projection layers. Second, because both the embeddings are in a common space, we compute the cosine similarity between each testing audio and all the class labels. Each audio will have as many logits as class labels. Third, logits are turned into a probability distribution by applying softmax for binary or multiclass and sigmoid for multilabel classification. 
\vspace{-0.15in}
\section{Experiments}\label{sec:experiments}
\vspace{-0.05in}
\begin{table*}
\small
\center
\adjustbox{max width=\textwidth}{%
\begin{tabular}{c|ccccc|cccc} \hline
& \multicolumn{5}{c|}{Sound Event Classification} & \multicolumn{3}{c}{Music} \\ \hline
Model & ESC50 & FSD50K & US8K & \makecell{DCASE17 \\ Task 4} & AudioSet  & \makecell{Music \\ Speech} & \makecell{Music \\ Genres} & \makecell{Mri. \\ Stroke} & \makecell{Mri. \\ Tonic}\\ \hline
Random & 0.02 &  $<$ 0.005 & 0.1 & 0.05 & $<$ 0.0018  & 0.5 & 0.1 & 0.1 & 0.1667 \\
Benchmark (ZS) & 0.6940\cite{audioclip} & 0.0302\cite{wav2clip} & 0.6531\cite{audioclip} & -  & - & - & - & - & - \\
CLAP(ZS) & \textbf{0.826} & \textbf{0.3024} & \textbf{0.7324} & \textbf{0.3} & \textbf{0.058}  & \textbf{1.0} & \textbf{0.252} & \textbf{0.3447} & \textbf{0.1965}\\ \hline
Benchmark (Best) & 0.9715 \cite{audioclip} & 0.641 \cite{turian2022hear} & 0.90-7 \cite{audioclip} & 0.646 \cite{kong2020sound} & 0.471 \cite{koutini2021efficient} & 0.992 \cite{turian2022hear} & 0.883 \cite{turian2022hear} & 0.975 \cite{turian2022hear} & 0.942 \cite{turian2022hear} \\
CLAP (Best) & 0.9670 & 0.5859 & 0.8796 & 0.5938 & -  & \textbf{1.0} & \textbf{0.9130} & \textbf{0.9794} & \textbf{0.9534}  \\ \hline
\end{tabular}
}
\smallskip
\center
\adjustbox{max width=\textwidth}{%
\begin{tabular}{c|c|c|cc|c|c|c} \hline
& \multicolumn{1}{c|}{\makecell{Instrument\\Classification}} & \multicolumn{1}{c|}{\makecell{Acoustic Scene \\ Classification}} & \multicolumn{2}{c|}{Emotion Recognition} & \multicolumn{1}{c|}{\makecell{Keyword \\ Spotting}} & \makecell{Vocal Sound \\ Classification} & \multicolumn{1}{c}{\makecell{Speaker \\ Counting}} \\ \hline
Model & \makecell{Beijing \\ Opera} & TUT2017 & \makecell{CRE\\MA-D} & \makecell{RAV\\DESS} & \makecell{Speech \\ Comm.} & \makecell{Vocal\\Sound} & \makecell{Libri \\ Count} \\ \hline
Random & 0.25 & 0.06 & 0.1667 & 0.125 & 0.083 & 0.1667 & 0.090 \\
CLAP (ZS) & \textbf{0.4746} & \textbf{0.2963} & \textbf{0.1784} & \textbf{0.1599} & \textbf{0.1063} & \textbf{0.4945} & \textbf{0.1788} \\ \hline
Benchmark (Best) & 0.975 \cite{turian2022hear} & 0.843 \cite{han2017convolutional} & 0.752 \cite{turian2022hear} & 0.8182 \cite{ravdess_best} & 0.987 \cite{speechcommands_best} & 0.905 \cite{vocalsound} & 0.785 \cite{turian2022hear} \\
CLAP (Best) & 0.9026 & 0.7463 & 0.6834 & 0.6436  & 0.9683 & \textbf{0.9795} & 0.7783 \\  \hline
\end{tabular}
}
\caption{\label{table: Zero-shot results}
CLAP (ZS) Zero-Shot outperforms the literature. CLAP (Best) is the best performance among our supervised setups. Higher is better for all numbers, DCASE17 employs F1, FSD50K and AudioSet employs mAP, everything else uses accuracy. \vspace{-0.1in}}
\end{table*}

\subsection{Datasets}
\vspace{-0.05in}

\noindent\textbf{Training.} We used 128,010 audio and text pairs from 4 datasets to construct the training dataset for CLAP. We extracted 36,796 pairs from FSD50k~\cite{fsd50k}, 29,646 pairs from ClothoV2 \cite{clotho}, 44,292 from AudioCaps \cite{audiocaps}, 17,276 pairs from MACS \cite{macs}. The dataset details are in appendix Section~\ref{appendix: training datasets} and Table~\ref{table: training dataset}.\\
\noindent\textbf{Downstream Tasks.} We used 16 datasets from 8 different domains as downstream tasks. Five tasks are Sound Event Classification. Five tasks are music related, classification of music vs speech, music genres, strokes and tonics. One task is Acoustic Scene Classification. Four are speech-based, Emotion Recognition, Keyword Spotting, and Vocal Sound Classification (e.g. cough, sneeze, laughter). One is counting speakers in a recording (0 to 10 speakers). The datasets are in Table~\ref{table: Zero-shot results}, and details are in appendix Section~\ref{appendix: downstream datasets} and Table~\ref{table: downstream datasets}. 
\vspace{-0.1in}
\subsection{Experimental setup}
\vspace{-0.05in}

\textbf{Pre-processing.} We used log Mel spectrogram representations of audio with a sampling rate of 44.1 KHz, hop size of 320 secs, window size 1024 secs, and 64 Mel bins in the range of 50-8000 Hz. During training, each audio clip is randomly truncated to a continuous segment of 5 secs, or padded if shorter. The captions were not altered. The batches with audio and text pairs are randomly sampled at training. \\
\textbf{Encoders.} We chose CNN14~\cite{pann} model as the audio encoder to provide a fair comparison to previous SoTA models. The model has 80.8 million parameters, an embedding size of 2048, and was pretrained with ~2M audio clips from AudioSet. The text encoder chosen is BERT \cite{devlin-etal-2019-bert}. We use HuggingFace \cite{wolf2019huggingface} implementation of BERT base uncased. The model has 110 million parameters. We limited the max text sequence length to 100 chars for computational efficiency. The [CLS] token from the final layer of BERT is used as the text embedding with a size of 768. Both, the audio and text embeddings are projected into a multimodal space with two learnable projection matrices resulting in an output dimension of 1024. The temperature parameter $\tau$ is learnable and initialised to 0.007. To prevent training instability, the logits scaled by $\tau$ are clipped to a maximum value of 100. \\
\textbf{Training.} We trained by unfreezing both encoders for 40 epochs. We use Adam Optimiser~\cite{adam} with an initial learning rate $10^{-3}$ and reduce the learning rate on plateau by $10^{-1}$ with a patience of 10. The models are implemented with PyTorch's Distributed Data-Parallel and use 16GB V100 GPUs with scaling from 8 to 24 GPUs.
\vspace{-0.2in}
\subsection{Evaluation setups for CLAP}
\vspace{-0.05in}

\noindent\textbf{Zero-shot Evaluation} studies the generalisation of CLAP to unseen classes and audios. The setup is explained in Section \ref{sec:method,subsec:zeroshot}. Instead of using the class label, we constructed a natural language prompting, \textit{`This is a sound of [class label]'}. The prompt was kept the same for all the domains except three. For Emotion Recognition we used \textit{`this person is feeling [class label]'}, for Keyword Spotting we only use the keyword, and for Speaker Counting we used \textit{`[number between 0 - 10] persons speaking'}. \\
\noindent\textbf{Supervised Feature Extraction Evaluation} studies the quality of audio representation learned by CLAP. Given a downstream task, we used CLAP as a feature extractor followed by training a classifier of 1 or 3 fully-connected layers (Freeze\_L1 and Freeze\_L3), similar to~\cite{pann}. We used a learning rate of $10^{-3}$ with Adam Optimizer for 30 epochs. We did not perform grid search for tuning hyperparameters due to computation constraints.\\ 
\noindent\textbf{Supervised Finetune Evaluation} benchmarks CLAP against the best performance for each task in the literature. Given a downstream task, we unfroze and finetuned the audio encoder together with an attached 1 or 3 fully-connected layers. We used a learning rate of $10^{-4}$ with Adam Optimizer for 30 epochs. We did not perform grid search for tuning hyperparameters due to computation constraints. \\
\vspace{-0.3in}

\section{Results and Discussion}\label{sec:results}
\vspace{-0.05in}
For baseline comparisons we considered the best model performances in the literature for Zero-Shot Learning `Benchmark (ZS)' and for Supervised Learning `Benchmark (Best)' and reported them in Table~\ref{table: Zero-shot results}. We also discuss the effect of freezing the encoders and the effect of prompts for Zero-Shot.

\vspace{-0.15in}
\subsection{Zero-Shot (ZS) results}\label{subsec:zero shot results}
\vspace{-0.05in}

CLAP (ZS) achieved SoTA on established Sound Event Classification (SEC) datasets like FSD50K, US8K and ESC50. For ESC50, CLAP achieved 82.6\% accuracy (acc) beating human performance of 81\% and AudioCLIP (69\%) by an absolute 12\%. In US8K, CLAP achieved 73\% acc outperforming AudioCLIP (65\%) by an absolute 8\%. For the multi-label dataset FSD50K, CLAP beat Wav2CLIP (3\%) by an absolute 27\% mAP. On task GTZAN's Music vs Speech Classification, CLAP even beat supervised models achieving 100\% acc. These results point to the possibility of having reliable audio models with no training involved.

CLAP (ZS) performed better than random on all downstream tasks and achieved good to slightly better than random on some music and speech-related tasks. CLAP achieved 47\% acc in Instrument Classification, an absolute 22\% higher than random. In the Vocal Sound dataset achieved 50\% acc, an absolute 33\% improvement over random. In Emotion Recognition (ER) and Keyword Spotting (KWS) CLAP outperformed random by up to an absolute 4\% acc.

\vspace{-0.15in}
\subsection{Supervised results}\label{subsec:finetune results}
\vspace{-0.05in}

CLAP (Best) is the best performance among supervised setups and achieved SoTA on 5 datasets. CLAP achieved in GTZAN Music vs Speech Classification 100\% acc, in GTZAN Music Genre Classification 91.3\% acc, in Mri. Stroke Classification 97.94\% acc, in Mri. Tonic Classification with 95.34\% acc, and in Vocal Sounds Classification 97.95\% acc. In other tasks CLAP underperformed SoTA by at most 7\%. The lowest performing task was ER's RAVDESS with 64\% acc vs a SoTA of 81\%.

CLAP performs better in domains like SEC than in others like ER, which was more evident in (ZS) than in (Best). We hypothesis that SEC tasks perform better because CLAP's training data consist of audio captioning datasets, which mainly include the description of sound events, acoustic scenes, actions, and objects. On the other hand, the training data is scarce on human speech and the captions do not describe aspects of its content or context. Therefore, CLAP underperforms on human speech tasks like KWS and ER. We posit that as we increase training data and increase human speech-based captioning, CLAP's performance on speech datasets will increase. 

\vspace{-0.15in}
\subsection{Effect of freezing CLAP encoders}
\vspace{-0.05in}
We studied how freezing the audio and/or text encoder during training affected performance of the downstream tasks. We computed the average of the CLAP (ZS) performance across the downstream tasks. The results are shown in \ref{table: frozen}. The best Avg. CLAP (ZS) score is obtained by unfreezing both encoders and the worst score by freezing both encoders. This is expected because unfreezing both encoders allows them to learn the multimodal information from the pairs. Surprisingly, unfreezing the text encoder performed better than unfreezing the audio encoder. Our intuition was that unfreezing the audio encoder would enable learning beyond the SEC coming from the pretrained AudioSet information. However, unfreezing the text encoder was better for CLAP, and a similar insight was found for CLIP models in Computer Vision~\cite{zhai2021lit}. This valuable finding suggests that, under the CLAP learning paradigm, it is possible to use an audio encoder of choice and turn it into a Zero-Shot classifier.

\begin{table}[h]
\small
\centering
\begin{tabular}{l|l|l|l} \hline
\makecell{Audio encoder \\ (frozen)} & \makecell{Text encoder \\ (frozen)} & \makecell{Avg. \\ ZS score} & \makecell{ESC50 \\ (acc)}\\ \hline
\cmark & \cmark & 0.2809 & 0.5555\\
\xmark &  \cmark &  0.2818 & 0.6415\\
 \cmark & \xmark &  0.3109 & 0.7631\\
\xmark & \xmark & 0.3265 &  0.826\\ \hline
\end{tabular}
\caption{\label{table: frozen}
Effect of freezing text and/or audio encoders on CLAP (ZS) performance across all tasks and ESC50.
}
\end{table}
\vspace{-0.3in}
\subsection{Changing prompts in Zero-Shot evaluation}
\vspace{-0.05in}
The training data of CLAP consists of natural language captions containing one or more sentences. However, the vast majority of datasets have class labels defined by a few words-- dog barking' and `sneezing'. Using single words instead of language description affects how Zero-Shot learning transfers. To overcome this distribution difference, we used standard template prompts \textit{`This is a sound of [class label]'}. Experimentally, we found that using appropriate prompts improved Zero-Shot performance of CLAP. For example, Table~\ref{table: prompt}, shows how changing prompts for ESC50 leads to a 5\% acc increase in performance. 

\begin{table}[h!]
\small
\centering
\begin{tabular}{l|l} \hline
\makecell{Prompt} & \makecell{ESC50 (acc)}\\ \hline
\textit{`i can hear [class label]'} & 0.786 \\
\textit{`this is an audio of [class label]'} & 0.8005 \\
\textit{`[class label]'} & 0.812 \\ 
\textit{`this is [class label]'} & 0.8135 \\
\textit{`this is a sound of [class label]'} & 0.826 \\
\hline
\end{tabular}
\caption{\label{table: prompt}
Effect of different prompts on ESC50 (ZS).
}
\end{table}

\vspace{-0.3in}

\section{Conclusion}\label{sec:conclusions}
\vspace{-0.05in}
We introduced CLAP\emojiclap for learning audio concepts from natural language supervision. CLAP does not require gold standard class labels for training, enables flexible class prediction, and generalizes to multiple downstream tasks. The training data consists of 128k audio-text pairs which is at least 0.001\% smaller than what similar Computer Vision models used. Nonetheless, CLAP establishes SoTA in Zero-Shot performance and SoTA in supervised performance for 5 tasks. Hence, CLAP shows potential for building an audio foundation model that can learn by natural language supervision and can generalize to a wide range of tasks while achieving SoTA performance.

\pagebreak

\bibliographystyle{IEEEbib}
\bibliography{refs}

\begin{thebibliography}{10}

\bibitem{lyon2010machine}
Richard~F Lyon,
\newblock ``Machine hearing: An emerging field [exploratory dsp],''
\newblock {\em IEEE signal processing magazine}, vol. 27, pp. 131--139, 2010.

\bibitem{mesaros2019sound}
Annamaria Mesaros, Aleksandr Diment, Benjamin Elizalde, Toni Heittola, Emmanuel
  Vincent, Bhiksha Raj, and Tuomas Virtanen,
\newblock ``Sound event detection in the dcase 2017 challenge,''
\newblock {\em IEEE/ACM Transactions on Audio, Speech, and Language
  Processing}, vol. 27, no. 6, pp. 992--1006, 2019.

\bibitem{zhang2021bigssl}
Yu~Zhang, Daniel~S Park, Wei Han, James Qin, Anmol Gulati, Joel Shor, Aren
  Jansen, Yuanzhong Xu, Yanping Huang, Shibo Wang, et~al.,
\newblock ``Bigssl: Exploring the frontier of large-scale semi-supervised
  learning for automatic speech recognition,''
\newblock {\em arXiv preprint arXiv:2109.13226}, 2021.

\bibitem{chen2021wavlm}
Sanyuan Chen, Chengyi Wang, Zhengyang Chen, Yu~Wu, Shujie Liu, Zhuo Chen, Jinyu
  Li, Naoyuki Kanda, Takuya Yoshioka, Xiong Xiao, et~al.,
\newblock ``Wavlm: Large-scale self-supervised pre-training for full stack
  speech processing,''
\newblock {\em arXiv preprint arXiv:2110.13900}, 2021.

\bibitem{baevski2020wav2vec}
Alexei Baevski, Yuhao Zhou, Abdelrahman Mohamed, and Michael Auli,
\newblock ``wav2vec 2.0: A framework for self-supervised learning of speech
  representations,''
\newblock {\em Advances in Neural Information Processing Systems}, vol. 33, pp.
  12449--12460, 2020.

\bibitem{radford2021learning}
Alec Radford, Jong~Wook Kim, Chris Hallacy, Aditya Ramesh, Gabriel Goh,
  Sandhini Agarwal, Girish Sastry, Amanda Askell, Pamela Mishkin, Jack Clark,
  et~al.,
\newblock ``Learning transferable visual models from natural language
  supervision,''
\newblock in {\em International Conference on Machine Learning}. PMLR, 2021,
  pp. 8748--8763.

\bibitem{yuan2021florence}
Lu~Yuan, Dongdong Chen, Yi-Ling Chen, Noel Codella, Xiyang Dai, Jianfeng Gao,
  Houdong Hu, Xuedong Huang, Boxin Li, Chunyuan Li, et~al.,
\newblock ``Florence: A new foundation model for computer vision,''
\newblock {\em arXiv preprint arXiv:2111.11432}, 2021.

\bibitem{jia2021scaling}
Chao Jia, Yinfei Yang, Ye~Xia, Yi-Ting Chen, Zarana Parekh, Hieu Pham, Quoc Le,
  Yun-Hsuan Sung, Zhen Li, and Tom Duerig,
\newblock ``Scaling up visual and vision-language representation learning with
  noisy text supervision,''
\newblock in {\em International Conference on Machine Learning}. PMLR, 2021,
  pp. 4904--4916.

\bibitem{wav2clip}
Ho-Hsiang Wu, Prem Seetharaman, Kundan Kumar, and Juan~Pablo Bello,
\newblock ``Wav2clip: Learning robust audio representations from clip,''
\newblock in {\em ICASSP 2022 - 2022 IEEE International Conference on
  Acoustics, Speech and Signal Processing (ICASSP)}, 2022, pp. 4563--4567.

\bibitem{audioclip}
Andrey Guzhov, Federico Raue, Jörn Hees, and Andreas Dengel,
\newblock ``Audioclip: Extending clip to image, text and audio,''
\newblock in {\em ICASSP 2022 - 2022 IEEE International Conference on
  Acoustics, Speech and Signal Processing (ICASSP)}, 2022, pp. 976--980.

\bibitem{audioSet}
Jort~F. Gemmeke, Daniel P.~W. Ellis, Dylan Freedman, Aren Jansen, Wade
  Lawrence, R.~Channing Moore, Manoj Plakal, and Marvin Ritter,
\newblock ``Audio set: An ontology and human-labeled dataset for audio
  events,''
\newblock in {\em 2017 IEEE International Conference on Acoustics, Speech and
  Signal Processing (ICASSP)}, 2017, pp. 776--780.

\bibitem{turian2022hear}
Joseph Turian, Jordie Shier, Humair~Raj Khan, Bhiksha Raj, Bj{\"o}rn~W
  Schuller, Christian~J Steinmetz, Colin Malloy, George Tzanetakis, Gissel
  Velarde, Kirk McNally, et~al.,
\newblock ``Hear 2021: Holistic evaluation of audio representations,''
\newblock {\em arXiv preprint arXiv:2203.03022}, 2022.

\bibitem{kong2020sound}
Qiuqiang Kong, Yong Xu, Wenwu Wang, and Mark~D Plumbley,
\newblock ``Sound event detection of weakly labelled data with cnn-transformer
  and automatic threshold optimization,''
\newblock {\em IEEE/ACM Transactions on Audio, Speech, and Language
  Processing}, vol. 28, pp. 2450--2460, 2020.

\bibitem{koutini2021efficient}
Khaled Koutini, Jan Schl{\"u}ter, Hamid Eghbal-zadeh, and Gerhard Widmer,
\newblock ``Efficient training of audio transformers with patchout,''
\newblock {\em arXiv preprint arXiv:2110.05069}, 2021.

\bibitem{han2017convolutional}
Yoonchang Han, Jeongsoo Park, and Kyogu Lee,
\newblock ``Convolutional neural networks with binaural representations and
  background subtraction for acoustic scene classification,''
\newblock {\em the Detection and Classification of Acoustic Scenes and Events
  (DCASE)}, pp. 1--5, 2017.

\bibitem{ravdess_best}
Cristina Luna-Jiménez, Ricardo Kleinlein, David Griol, Zoraida Callejas,
  Juan~M. Montero, and Fernando Fernández-Martínez,
\newblock ``A proposal for multimodal emotion recognition using aural
  transformers and action units on ravdess dataset,''
\newblock {\em Applied Sciences}, vol. 12, no. 1, 2022.

\bibitem{speechcommands_best}
Byeonggeun Kim, Simyung Chang, Jinkyu Lee, and Dooyong Sung,
\newblock ``Broadcasted residual learning for efficient keyword spotting,''
\newblock {\em arXiv preprint arXiv:2106.04140}, 2021.

\bibitem{vocalsound}
Yuan Gong, Jin Yu, and James Glass,
\newblock ``Vocalsound: A dataset for improving human vocal sounds
  recognition,''
\newblock in {\em ICASSP 2022 - 2022 IEEE International Conference on
  Acoustics, Speech and Signal Processing (ICASSP)}, 2022, pp. 151--155.

\bibitem{fsd50k}
Eduardo Fonseca, Xavier Favory, Jordi Pons, Frederic Font, and Xavier Serra,
\newblock ``Fsd50k: An open dataset of human-labeled sound events,''
\newblock {\em IEEE/ACM Transactions on Audio, Speech, and Language
  Processing}, vol. 30, pp. 829--852, 2022.

\bibitem{clotho}
Konstantinos Drossos, Samuel Lipping, and Tuomas Virtanen,
\newblock ``Clotho: an audio captioning dataset,''
\newblock in {\em ICASSP 2020 - 2020 IEEE International Conference on
  Acoustics, Speech and Signal Processing (ICASSP)}, 2020, pp. 736--740.

\bibitem{audiocaps}
Chris~Dongjoo Kim, Byeongchang Kim, Hyunmin Lee, and Gunhee Kim,
\newblock ``{AudioCaps: Generating Captions for Audios in The Wild},''
\newblock in {\em NAACL-HLT}, 2019.

\bibitem{macs}
Irene Mart{\'\i}n-Morat{\'o} and Annamaria Mesaros,
\newblock ``What is the ground truth? reliability of multi-annotator data for
  audio tagging,''
\newblock in {\em 2021 29th European Signal Processing Conference (EUSIPCO)}.
  IEEE, 2021, pp. 76--80.

\bibitem{pann}
Qiuqiang Kong, Yin Cao, Turab Iqbal, Yuxuan Wang, Wenwu Wang, and Mark~D.
  Plumbley,
\newblock ``Panns: Large-scale pretrained audio neural networks for audio
  pattern recognition,''
\newblock {\em IEEE/ACM Trans. Audio, Speech and Lang. Proc.}, vol. 28, pp.
  2880–2894, jan 2020.

\bibitem{devlin-etal-2019-bert}
Jacob Devlin, Ming-Wei Chang, Kenton Lee, and Kristina Toutanova,
\newblock ``{BERT}: Pre-training of deep bidirectional transformers for
  language understanding,''
\newblock in {\em Proceedings of the 2019 Conference of the North {A}merican
  Chapter of the Association for Computational Linguistics: Human Language
  Technologies, Volume 1 (Long and Short Papers)}, Minneapolis, Minnesota, June
  2019, pp. 4171--4186, Association for Computational Linguistics.

\bibitem{wolf2019huggingface}
Thomas Wolf, Lysandre Debut, Victor Sanh, Julien Chaumond, Clement Delangue,
  Anthony Moi, Pierric Cistac, Tim Rault, R{\'e}mi Louf, Morgan Funtowicz,
  et~al.,
\newblock ``Huggingface's transformers: State-of-the-art natural language
  processing,''
\newblock {\em arXiv preprint arXiv:1910.03771}, 2019.

\bibitem{adam}
Diederik~P. Kingma and Jimmy Ba,
\newblock ``Adam: A method for stochastic optimization,''
\newblock in {\em ICLR (Poster)}, 2015.

\bibitem{zhai2021lit}
Xiaohua Zhai, Xiao Wang, Basil Mustafa, Andreas Steiner, Daniel Keysers,
  Alexander Kolesnikov, and Lucas Beyer,
\newblock ``Lit: Zero-shot transfer with locked-image text tuning,''
\newblock {\em arXiv preprint arXiv:2111.07991}, 2021.

\end{thebibliography}

\clearpage
\appendix

\section{Training datasets} \label{appendix: training datasets}

\noindent \textbf{FSD50k}~\cite{fsd50k} is a sound event classification dataset with audio clips from freesound.org. The duration of the clips ranges from 0.3 to 30 seconds. We used the ~36k clips from training and validation. We constructed the caption for each clip by concatenating the two sentences the associated title and description in the metadata. We ignored the class label. \\
\textbf{ClothoV2} \cite{clotho} is an audio captioning dataset consisting of ~7k audio clips. The duration of the clips range from 15 to 30 seconds. Each clip has 5 captions annotated by different participants. Thus, we created 5 pairs for each clip extending the number of audio-text pairs by 5 times. \\
\textbf{AudioCaps} \cite{audiocaps} is an audio captioning dataset consisting of ~46k audio clips from AudioSet. The duration of the clips is 10 seconds. Each clip has a caption annotated via crowd-sourcing. \\
\textbf{MACS} \cite{macs} is an audio captioning dataset consisting of ~4k audio clips. The duration of the clips is 10 seconds. Each clip is captioned by multiple participants. Similar to ClohtoV2, we paired the same audio with a each of their associated captions to create a larger set of pairs consisting of ~17k. 
At the time of downloading the datasets, not all clips were available from the web links.

\begin{table}[ht]
\center
\begin{tabular}{lccc} \hline
Dataset & Pairs & \makecell{Unique\\ audios} & \makecell{Unique \\captions} \\ \hline
FSD50k & 36,796 & 36,796 & 36,796 \\
ClothoV2 & 29,646 & 5,929 & 29,646 \\
AudioCaps & 44,292 & 44,292 & 44,292 \\
MACS & 17,276 & 3,930 & 17,276 \\ \hline
 & 128,010 & 90,947 & 128,010 \\ \hline
\end{tabular}
\caption{\label{table: training dataset}
Training dataset statistics. \vspace{-0.08in}}
\end{table}

\begin{table*}[ht]
\small
\center
\begin{tabular}{ccccccccc}\hline
 Domain & Dataset & Files & Dur. (secs) & Classes & Metric & Setup \\ \hline
\multirowcell{5}{
 Sound Event  \\ Classification (SEC)} & ESC50 & 2k & 5 & 50 & ACC & 5 folds \\
 & FSD50K & ~51k & 0.3 - 30 & 200 & mAP & train/val/test \\
 & UrbanSound8K & ~8k & $\leq$ 4 & 10 & ACC & 10 folds \\
 & DCASE2017 Task4 & 52k & 10 & 17 & ACC & train/val/test \\
 & AudioSet & $\sim$2M & 10 & 527 & mAP & train/val/test \\ \hline
\multirowcell{5}{Music} & GTZAN Music Speech & 120 & 30 & 2 & ACC & 10 folds \\
 & GTZAN Music Genre & 1k & 30 & 10 & ACC & 10 folds \\
 & Mridangam Stroke & ~7k & 0.81 & 10 & ACC & 5 folds \\
 & Mridangam Tonic & ~7k & 0.81 & 6 & ACC & 5 folds \\ \hline
\makecell{Instrument \\ Classification} & \makecell{Beijing Opera \\ Percussions} & 236 & 4.77 & 4 & ACC & 5 folds \\ \hline
\makecell{Acoustic Scene \\ Classification} & TUT 2017 & 6.3k & 10 & 15 & ACC & train/val/test \\ \hline
\multirowcell{2}{Emotion \\ Recognition} & CREMA-D & ~7k & 5 & 6 & ACC & 5 folds \\
 & RAVDESS & ~2.5k & $\leq$ 5 & 8 & ACC & 5 folds \\ \hline
Keyword \\ Spotting & Speech Commands & 100k & 1 & 12 & ACC & train/val/test \\ \hline
\makecell{Vocal Sound \\ Classification} & \makecell{Vocal Sound} & ~21k & 5 & 6 & ACC & train/val/test \\ \hline
Speaker Counting & LibriCount 10 & 5k & 5 &  11 & ACC & 5 folds \\\hline
\end{tabular}
\caption{\label{table: downstream datasets}
Details from the 16 datasets used as Downstream Tasks.
}
\end{table*}

\section{Downstream datasets} \label{appendix: downstream datasets}
\vspace{-0.05in}
\textbf{ESC50} is an environmental classification dataset comprising of 50 events. The dataset consists of 2k files of 5 seconds each. The evaluation setup is 5 fold cross validation and the evaluation metric is accuracy. \\
\textbf{FSD50K} is a sound event classification dataset comprising of 200 events. The dataset consists of 51k files ranging from 0.3 to 30 seconds each. The evaluation setup is train/val/test and the evaluation metric is mAP. \\
\textbf{UrbanSound8K} is urban sound classification dataset comprising of 10 sounds. The dataset consists of 8k files of ~4 seconds each. The evaluation setup is 10 fold cross validation and the evaluation metric is accuracy. \\
\textbf{DCASE2017 Task4} is a sound event classification dataset comprising of 17 sounds recorded in domestic environment. The dataset consists of ~30k files of 10 seconds each. The evaluation setup is train/val/test and the evaluation metric is accuracy. \\
\textbf{AudioSet} is a sound event classification dataset comprising of 527 sounds from YouTube videos. The dataset consists of ~2M files of 10 seconds each. The evaluation setup is train/val/test and the evaluation metric is accuracy. \\
\textbf{TUT 2017} is an acoustic scene classification dataset comprising of 15 acoustic scenes in both outdoor and indoor environment. The dataset consists of ~52k files of 10 seconds each. The evaluation setup is train/val/test and the evaluation metric is accuracy.\\ 
\textbf{GTZAN Music Speech} is a binary classification dataset where the aim is to distinguish between human speech and music. The dataset consists of 120 files of 30 seconds each. The evaluation setup is 10 fold cross validation and the evaluation metric is accuracy. \\
\textbf{GTZAN Genres} is music genre classification dataset comprising of 10 genres. The dataset consists of 1k files of 30 seconds each. The evaluation setup is 10 fold cross validation and the evaluation metric is accuracy. \\
\textbf{Mridangam Stroke} is music stroke classification dataset comprising of 10 strokes from Mridangam (pitched percussion instrument). The dataset consists of 1k files of 0.81 seconds each. The evaluation setup is 5 fold cross validation and the evaluation metric is accuracy. \\
\textbf{Mridangam Tonic} is music tonic classification dataset comprising of 6 tonics from Mridangam (pitched percussion instrument). The dataset consists of 1k files of 0.81 seconds each. The evaluation setup is 5 fold cross validation and the evaluation metric is accuracy. \\
\textbf{Beijing Opera Percussions} is an instrument classification dataset comprising of 4 percussion instruments from Beijing Opera. The dataset consists of 236 files of 4.77 seconds each. The evaluation setup is 5 fold cross validation and the evaluation metric is accuracy. \\
\textbf{CREMA-D} is an emotion recognition dataset comprising of 6 emotions. The dataset consists of ~7k files of 5 seconds each. The evaluation setup is 5 fold cross validation and the evaluation metric is accuracy. \\
\textbf{RAVDESS} is an emotion recognition dataset comprising of 8 emotions. The dataset consists of ~2.5k files of 5 seconds each. The evaluation setup is 5 fold cross validation and the evaluation metric is accuracy. \\
\textbf{Speech Commands V2} is an keyword spotting dataset comprising of 13 commands. The dataset consists of 100k files of 1 seconds each. The evaluation setup is train/val/test and the evaluation metric is accuracy.\\
\textbf{Vocal Sound} is a human vocal sound classification dataset comprising of 6 vocalizations. The dataset consists of 21k files of 5 seconds each. The evaluation setup is train/val/test and the evaluation metric is accuracy.\\
\textbf{LibriCount} is a speaker count estimation dataset comprising of simulated cocktail party environment audios consisting of 0 to 10 speakers. The dataset consists of 5k files of 5 seconds each. The evaluation setup is 5 fold cross validation and the evaluation metric is accuracy. \\

\begin{table*}[ht]
\small
\center
\begin{tabular}{c|cccc|cccc} \hline
\small
 & \multicolumn{4}{c|}{Sound Event Classification} & \multicolumn{4}{c}{Music} \\\hline
Model & ESC50 & FSD50K & US8K & \makecell{DCASE17 \\ Task 4} & \makecell{Music \\ Speech} & \makecell{Music \\ Genres} & \makecell{Mri. \\ Stroke} & \makecell{Mri. \\ Tonic}\\ \hline
YAMNet & 0.8375 & - & - & -   & 0.969 & 0.847 & - & - \\
Open L3 & 0.7505 & 0.4470 & 0.7823 & -   & 0.969 & 0.879 & 0.9666 & 0.9369 \\
Wav2CLIP & 0.7589 & 0.3617 & - & -   & 0.946 & 0.748 & 0.9471 & 0.8289 \\
PaNN & 0.9085 & - & - & -   & 0.992 & 0.860 & 0.9390 & 0.8244 \\
Wav2Vec2 & 0.5610 & 0.1164 & - & -   & 0.946 & 0.780 & 0.9432 & 0.8283 \\ 
CLAP (S) & 0.9310 & 0.5905 & 0.8389 & 0.5330  & 1.0 & 0.7930 & 0.7754 & 0.6391 \\
CLAP (F) & 0.9670 & 0.5859 & 0.8796 & 0.5938  & 1.0 & 0.9130 & 0.9794 & 0.9534  \\ \hline
\end{tabular}
\smallskip
\center
\begin{tabular}{c|c|c|cc|c|c|c} \hline
\small
& \multicolumn{1}{c|}{\makecell{Instrument\\Classification}} & \multicolumn{1}{c|}{\makecell{Acoustic Scene\\Classification}} & \multicolumn{2}{c|}{Emotion Recognition} & \multicolumn{1}{c|}{\makecell{Keyword \\ Spotting}} & \makecell{Human \\ Vocalization} & \multicolumn{1}{c}{\makecell{Speaker \\ Estimation}} \\ \hline
Model & \makecell{Beijing \\ Opera} & \makecell{TUT 2017} & \makecell{CRE\\MA-D} & \makecell{RAV\\DESS} &  \makecell{Speech \\ Comm.} & \makecell{Vocal\\Sounds} & \makecell{Libri \\ Count} \\ \hline
YAMNet & 0.941 & - & 0.453  & 0.479 & 0.4104 & - & 0.6526 \\
OpenL3 & 0.975 & - & 0.550  & 0.604 & 0.7634 & - & 0.6414 \\
Wav2CLIP & 0.936 & - & 0.512  & 0.684 & 0.3466 & - & 0.5276 \\
PaNN & 0.911 & - & 0.555  & 0.429 & 0.6182 & - & 0.6516 \\
Wav2Vec2 & 0.907 & - & 0.6562  & - & 0.8785 & - & 0.6921 \\
CLAP (S) & 0.7754 & 0.7099 & 0.2830 & 0.4515  & 0.3708 & 0.8411 & 0.5715 \\
CLAP (F) & 0.9026 & 0.7463 & 0.6834 & 0.6436  & 0.9683 & 0.9795 & 0.7783 \\ \hline
\end{tabular}
\caption{\label{table: shallow results}
CLAP's Supervised Feature Extraction (S) and Supervised Finetune (F) performance against SoTA models.
}
\end{table*}

\begin{table*}[ht]
\small
\center
\begin{tabular}{lll} \hline
Dataset & Audio captions \\ \hline
ClothoV2 & A bow playing a stringed instrument in a one note tone repeatedly before violins join to create the melody\\
ClothoV2 & An insect buzzing in the foreground as birds chirp in the background\\
ClothoV2  & A camp fire crackles as the flames burn branches and leaves \\
\hline
AudioCaps & Several sirens are wailing and a horn is honked twice \\
AudioCaps & Church bells chime loudly and repeatedly\\
AudioCaps & Beating drum getting faster than children voices and clapping then adult male voice \\ \hline
FSD50K & Water dripping in a cave or underground temple \\
FSD50K & Canada geese flying down and landing near a lakeshore. Recorded with an Olympus LS-14. \\
FSD50K & Leaves falling in a forest near a pond. Recorded in October 2017 in a German forest using a Zoom H2n. \\ \hline
MACS & Two people having a conversation nearby while a lot of adults  and a child talk far away \\
MACS & Birds are making a lot of noises and a distant child yells \\
MACS & Dog barks followed by adults talking and children voices \\ \hline
\end{tabular}
\caption{\label{table: caption example}
Randomly sampled raw captions from each dataset.
}
\end{table*}

\section{Discussion} \label{appendix: discussion}

\subsection{Batch size and CLAP performance}

Appropriate batch size is a key contributor in performance of any contrastive learning methods that uses positive-negative pairs. In computer vision literature, larger batch size has shown to improve performance \cite{radford2021learning, yuan2021florence, zhai2021lit}. 

We use Tesla V100 16 GB GPUs for this analysis. The number of GPUs used vary from 4 to 24 for batch size from 32 to 768. We measure the zero-shot performance by computing average performance across N downstream tasks listed in table \ref{table: downstream datasets}. Our findings also indicate that the larger batch size does lead to improved CLAP performance. However, we saw decreased performance with batch size of 768. This might be an anamoly in an increasing zero-shot performance trend. We leave the larger batch size investigation to future work. 

\begin{figure}[ht]
  \centering
     \includegraphics[width=2.5in]{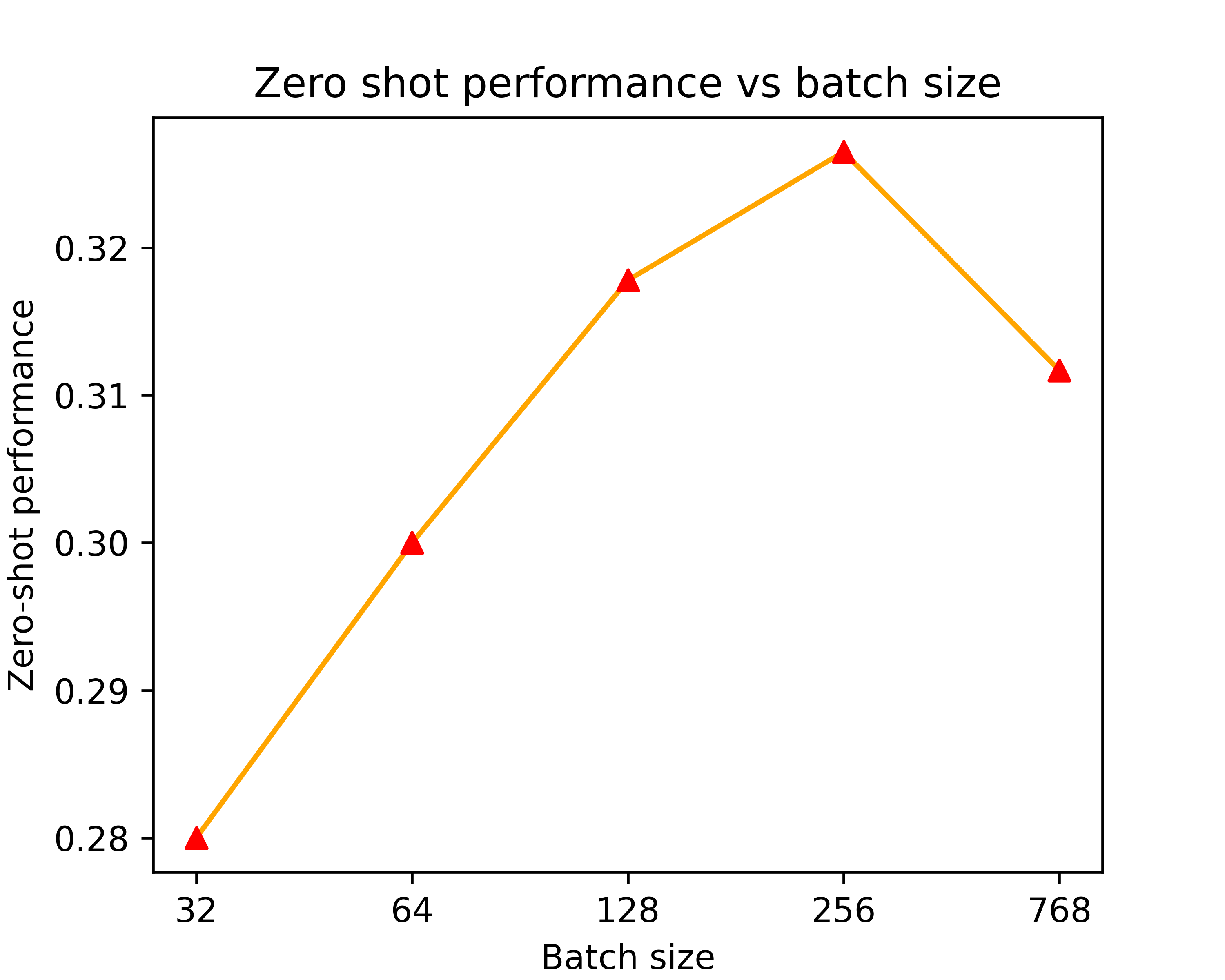}
     \caption{Effect of batch size on zero-shot performance}
     \label{fig:clap_batch_size}
\end{figure}

\subsection{Training with AudioSet}
AudioSet is a multilabled sound event dataset with ~2M audio clips, but it does not come with a text descriptions per clip, thus making it complicated to extract audio and text pairs for training. We tried constructing the text description with the title and the class label(s). However, adding about ~1.7M pairs to the existing 128k pairs resulted in a performance dropped in the overall zero-shot numbers. For example, ESC50 performance dropped from 82.6\% to 67.15\% acc and US8K dropped from 73.24\% to 70.93\% acc. Only Speech Commands V2 (SCV2) performance improved from 10\% to 15\% acc. Perhaps due to the large amount of audio containing speech in AudioSet. The quality of audio-text pairs is key in training CLAP. In AudioSet, often the YouTube titles and descriptions do not describe the acoustic content of the video segment under consideration but instead describe the video as a whole. More intelligent ways of generating descriptions can benefit CLAP training with AudioSet and other many similar datasets. In general, finding helpful training data for CLAP based on public datasets is difficult, thus relying on large-scale noisy pairs is the only scalable approach.

\section{Acknowledgements} \label{appendix: acknowledgements}
We thank Hamid Eghbalzadeh for early discussions and feedback.

\end{document}